\documentclass[aps,10pt]{revtex4}
\usepackage{rotating}
\usepackage{amssymb}
\usepackage{amsmath}
\usepackage{amsfonts}
\usepackage{graphicx}
\usepackage{graphics}
\usepackage{psfig}

\begin{document}

\title{Gauge Theories with Lorentz-Symmetry Violation by Symplectic
Projector Method}
\author{H. Belich}
\email{belich@cce.ufes.br}
\affiliation{{\small {Universidade Federal do Esp\'{\i}rito Santo (UFES), Departamento de
F\'{\i}sica e Qu\'{\i}mica, Av. Fernando Ferrari, S/N Goiabeiras, Vit\'{o}%
ria - ES, 29060-900 - Brasil}}}
\affiliation{{\small {Grupo de F\'{\i}sica Te\'orica Jos\'e Leite Lopes, C.P. 91933, CEP
25685-970, Petr\'opolis, RJ, Brasil}}}
\author{M. A. De Andrade}
\email{marco@cbpf.br}
\affiliation{{\small {Universidade do Estado do Rio de Janeiro (UERJ), Departamento de
Mec\^anica e Energia, Estrada Resende Riachuelo s/n - Morada da Colina, CEP
27523-000, Resende - RJ}}}
\affiliation{{\small {Grupo de F\'{\i}sica Te\'orica Jos\'e Leite Lopes, C.P. 91933, CEP
25685-970, Petr\'opolis, RJ, Brasil}}}
\author{M. A. Santos}
\email{masantos@cce.ufes.br}
\affiliation{{\small {Universidade Federal do Esp\'{\i}rito Santo (UFES), Departamento de
F\'{\i}sica e Qu\'{\i}mica, Av. Fernando Ferrari, S/N Goiabeiras, Vit\'{o}%
ria - ES, 29060-900 - Brasil}}}

\begin{abstract}
The violation of Lorentz symmetry is studied from the point of view of a
canonical formulation. We make the usual analysis on the constraints
structure of the Carroll-Field-Jackiw model. In this context we
derive the equations of motion for the physical variables and check out the
dispersion relations obtained from them. Therefore, by the analysis
using Symplectic Projector Method (SPM), we can check the results about
this type of Lorentz breaking with those in the recent literature: in this sense 
we can confirm that the configuration of 
$v^{\mu }$ space-like is stable, and the $v^{\mu }$ time-like carry
tachionic \ modes.
\end{abstract}

\maketitle

\section{ Introduction}

Investigations about repercussions of spontaneous symmetry breaking in
context of a fundamental theory has gain special attention in the last
years. The immediate consequence is the non equivalence between particle and
observer Lorentz transformations \cite{observ} . In the last decade some works 
\cite{Kostelec1} explored this breaking in the context of string theories.
These models with Lorentz and CPT breaking were used as a low-energy limit
of an extension of the standard model, valid at the Plank scale \cite%
{Colladay}. An effective action that incorporates CPT and Lorentz violation
is obtained and it keeps unaffected the $SU(3)\times SU(2)\times U(1)$ gauge
structure of the underlying theory. In the context of $N=1$ - supersymmetric
models, there have appeared two proposals: the one which violates the
algebra of supersymmetry was first addressed by Berger and Kostelecky \cite%
{Berger}, and other that preserve the SUSY algebra gives,after integration
on Grassman variables, the Carroll-Field-Jackiw model \cite{nos}. More
recently, in the work of ref. \cite{NIBBELINK}, the authors addressed the
discussion of Lorentz symmetry breaking in the context of exact SUSY by
working out a whole list of manifestly supersymmetric Lorentz-violating
operators. The observation of ultra-high energy cosmic rays with energies
beyond the Greisen-Zatsepin-Kuzmin (GZK) cutoff $\left( E_{GZK}\simeq 4\cdot
10^{19}\,eV\right) $ \cite{Cosmic}, \cite{Coleman}, could be potentially
taken to be as an evidence of Lorentz-violation. The rich phenomenology of
fundamental particles has also been considered as a natural environment in
the search for indications of breaking of these symmetries \cite{Coleman},%
\cite{Particles}, revealing up the moment stringent limitations on the
factors associated with such violation. Another point of interest refers to
ancient issue of space-time varying coupling constants \cite{Coupling},
which has been reassessed in the light of Lorentz-violating theories, with
interesting connections with the construction of supergravity models.
Moreover, measurements of radio emission from distant galaxies and quasars
put in evidence that the polarization vectors of the radiation emitted are
not randomly oriented as naturally expected. This peculiar phenomenon
suggests that the space-time intervening between the source and observer may
 exhibit some sort of optical activity (birefringence), whose origin is
unknown \cite{Astrophys}. There are some different proposals of the Lorentz
violation: one of these consists in obtain this breaking from spontaneous
symmetry breaking of a vector matter field \cite{colatto}.

Our approach to the Lorentz breaking consist in adopting the 4-dimensional
version of a Chern-Simons topological term, namely $\epsilon _{\mu \nu
\kappa \lambda }v^{\mu }A^{\nu }F^{\kappa \lambda }$, where $\epsilon _{\mu
\nu \kappa \lambda }$ is the 4-dimensional Levi-Civita symbol and $v^{\mu }$
is a fixed four-vector acting as a background. This idea was first settled
down in the context of QED in \cite{Jackiw}. A study of the consequences of
such breaking in QED is extensively analyzed in \cite{Adam}, \cite{Chung}.
An extension of the Carroll-Field-Jackiw model in $\left( 1+3\right) $
dimensions, including a scalar sector that yields spontaneous symmetry
breaking (Higgs sector), was recently developed and analyzed, resulting in
an Abelian-Higgs gauge model with violation of Lorentz symmetry \cite{Belich}%
.

The dimensional reduction (to 1+2 dimensions) of the Carroll-Field-Jackiw
model was successfully realized \cite{Manojr1}, \cite{Manojr2} , resulting
in a planar theory composed by a Maxwell-Chern-Simons gauge field, a
massless scalar field, and a mixing term (responsible by the Lorentz
violation) that couples both of these fields.

In this paper we explore the physical content of this model from the
viewpoint of a canonical formulation; incidentally, as far as we know, it
has not yet been treated in the literature. So, in section II, we make the
usual analysis on the constraints structure of the model. There we get
restrictions on the vector $v^{\mu }$ still as a condition to have a
consistent constrained (or gauge) theory. Next, in section III, we apply the
so-called Symplectic Projector Method (SPM), a recently developed
alternative to Dirac's most traditional methods, in which the physical
viewpoint is most hardly lost. In this context we derive the equations of
motion for the physical variables and check out the dispersion relations
obtained here with that in \cite{Jackiw}. Finally, in section IV we draw our
conclusions and make some general remarks.

\section{The Maxwell - Chern-Simons Model}

One starts from the Maxwell Lagrangian in 1+3 dimensions supplemented with a
term that couples the dual electromagnetic tensor to a fixed 4-vector, $%
v^{\mu },$ as it appears in ref. \cite{Belich}: 
\begin{equation}
\mathcal{L}_{1+3}=-\frac{1}{4}F^{\mu \nu }F_{\mu \nu }-\frac{1}{4}%
\varepsilon ^{\mu \nu \kappa \lambda }v_{\mu }A_{\nu }F_{\kappa \lambda }.
\label{action1}
\end{equation}
This model, as discussed in\cite{Jackiw}, is gauge invariant but does not
preserve the Lorentz and CTP symmetries. It is interesting to remark that
this four-vector $v_{\mu }$\ is rigid when we apply a rotation or generic
observable Lorentz 's boost. The phase velocity of light changes and we have
two modes of propagation moving with different velocities. As consequence we
have a rotation of the polarization plane (vacuum birefringence). The change
in the wave equation by the Chern-Simons term also opens the possibility of
obtain imaginary frequencies. These unstable modes can be avoided
establishing condition in our four-vector. In the last section we analyze
the dispersion relation of the model and return to these questions. Now we
are going to make another analysis under the canonical formalism. Taking as
starting point, we assume no restriction on the four-vector $v_{\mu }$\. The
restriction \textbf{\ }$v_{\mu }$ constant will be obtained as consistence
condition of the model viewed as constrained theory.

\subsection{The Constraints Analysis}

The lagrangian \ref{action1} carries the primary constraint 
\begin{equation}
\Phi _{1}=\pi ^{0}\approx 0.  \label{v1}
\end{equation}%
Together with the the primary Hamiltonian 
\begin{equation}
\mathcal{H}_{p}=\frac{1}{2}\pi ^{i}\pi ^{i}+\frac{1}{4}F^{ij}F^{ij}-\frac{1}{%
2}\varepsilon _{0ijk}\left( \pi ^{i}v^{j}A^{k}-v^{0}A^{k}\partial
^{i}A^{j}\right) +\frac{1}{8}\left(
v_{j}v^{j}A_{k}A^{k}-v_{j}A^{j}v_{k}A^{k}\right) +A^{0}\left( \partial
_{i}\pi _{i}-\frac{1}{2}\varepsilon _{0ijk}v^{k}\partial ^{i}A^{j}\right) ,
\label{3}
\end{equation}%
the consistency condition on eq.\ref{v1}, 
\begin{equation}
\left\{ \Phi _{1},H_{p}\right\} \approx 0,  \label{4}
\end{equation}%
generates the secondary constraint 
\begin{equation}
\Phi _{2}=\partial _{i}\pi _{i}-\frac{1}{2}\varepsilon _{0ijk}v^{k}\partial
^{i}A^{j}\approx 0.  \label{v2}
\end{equation}%
Now, analyzing the consistency condition on eq. \ref{v2}, we conclude that
the Dirac's algorithm turns out to produce a restriction on $v^{\mu }$ , 
\begin{equation}
\varepsilon ^{0ijk}\pi _{i}\partial _{k}v_{j}-\varepsilon ^{0ijk}\partial
_{i}A_{j}\partial _{k}v_{0}-\frac{1}{4}A_{k}\partial
^{k}v^{2}-v_{j}A^{k}\partial ^{j}v_{k}=0,  \label{6}
\end{equation}%
that must be preserved in order that a consistent constraint theory can be
built So, with the stronger restriction 
\begin{equation}
\partial_{\mu}v^{\mu}=0,  \label{7}
\end{equation}%
as made by Jackiw, we have a set of two, first-class, constraints (eq. \ref%
{v1} and \ref{v2}), and the canonical Hamiltonian 
\begin{equation}
\mathcal{H}=\frac{1}{2}\pi ^{i}\pi ^{i}+\frac{1}{4}F^{ij}F^{ij}-\frac{1}{2}%
\varepsilon _{0ijk}\left( \pi ^{i}v^{j}A^{k}-v^{0}A^{k}\partial
^{i}A^{j}\right) +\frac{1}{8}\left(
v_{j}v^{j}A_{k}A^{k}-v_{j}A^{j}v_{k}A^{k}\right).  \label{Ham1}
\end{equation}%
As \ we are going to apply the SPM, a consistent set of gauge-fixing
conditions must be imposed. We propose to work on the radiation conditions: 
\begin{eqnarray}
\Phi _{3} &=&A_{0}\approx 0,  \label{gauge1} \\
\Phi _{4} &=&\partial ^{i}A_{i}\approx 0.  \notag
\end{eqnarray}%
So the counting of degrees of freedom indicates that, after applying the
SPM, we can expect a 4-dimensional reduced, or physical, phase-space.

\section{The Symplectic Projector Method}

Basically, two approaches are possible to treat problems with gauge
symmetries in the canonical framework, since the presence of spurious
coordinates is the mark of such theories: one can either reduce or expand
the original space \ where act the constraints. Although the latter is the
choice where the problem have been treated in a more complete and successful
way (\cite{BRST}), the first one may offer, at least in simpler models, an
alternative where neither strong mathematical tools are needed nor the
physical viewpoint may be lost. In this sense, in recent years, the SPM have
been developed and applied to several multidimensional problems ( \cite%
{Amaral1}, \cite{Marco2, Marco3, Marco4, Marco5, Marco6}). However, since
this is not a very widespread method, we report to (\cite{Marco3}), where a
recent and brief review has been made on it's construction, so as we can
avoid here a very exhaustive \ paper.

The matrix whose elements are $g^{ij}\equiv \left\{ \Phi ^{i},\Phi
^{j}\right\} $ is easily calculated, and it's inverse , $g^{-1}\left(
x,y\right) $ is 
\begin{equation}
g^{-1}=\left[ 
\begin{array}{cccc}
0 & 0 & \delta ^{3}\left( x-y\right) & 0 \\ 
0 & 0 & 0 & \nabla ^{-2} \\ 
-\delta ^{3}\left( x-y\right) & 0 & 0 & 0 \\ 
0 & -\nabla ^{-2} & 0 & 0%
\end{array}%
\right].  \label{M2}
\end{equation}%
The general form of the symplectic projector is given by (\cite{Marco3}) 
\begin{equation}
\Lambda _{\nu }^{\mu }\left( x,y\right) =\delta _{\nu }^{\mu }\delta
^{3}\left( x-y\right) -\epsilon ^{\mu \alpha }\int d^{3}\tau d^{3}\varpi
g_{ij}\left( \tau ,\varpi \right) \delta _{\alpha (x)}\Phi ^{i}\left( \tau
\right) \delta _{\nu (y)}\Phi ^{j}\left( \varpi \right),  \label{c2}
\end{equation}%
with $\delta _{\alpha (x)}\Phi ^{i}\left( \tau \right) \equiv \frac{\delta
\Phi ^{i}\left( \tau \right) }{\delta \xi ^{\alpha }(x)}$ and $\epsilon
^{\mu \alpha }$ is the symplectic matrix element. Let's make the following
correspondence: 
\begin{equation}
\left( A^{0},A^{1},A^{2},A^{3},\pi _{0},\pi _{1},\pi _{2},\pi _{3}\right)
\Longleftrightarrow \left( \xi _{1},\xi _{2},\xi _{3},\xi _{4},\xi _{5},\xi
_{6},\xi _{7},\xi _{8}\right) .  \label{c3}
\end{equation}%
After some long and straightforward calculations, we find the projector matrix below. 
\begin{sidewaystable}
\begin{equation*}
\Lambda =  \label{M3}
\left[ 
\begin{array}{cccccccc}
0 & 0 & 0 & 0 & 0 & 0
& 0 & 0 \\ 
0 & \delta -\partial _{1}^{x}\partial _{1}^{y}\nabla ^{-2} & -\partial
_{1}^{x}\partial _{2}^{y}\nabla ^{-2} & -\partial _{1}^{x}\partial
_{3}^{y}\nabla ^{-2} & 0 & 0 & 0 & 0 \\ 
0 & -\partial _{2}^{x}\partial _{1}^{y}\nabla ^{-2} & \delta -\partial
_{2}^{x}\partial _{2}^{y}\nabla ^{-2} & -\partial _{2}^{x}\partial
_{3}^{y}\nabla ^{-2} & 0 & 0 & 0 & 0 \\ 
0 & -\partial _{3}^{x}\partial _{1}^{y}\nabla ^{-2} & -\partial
_{3}^{x}\partial _{2}^{y}\nabla ^{-2} & \delta -\partial _{3}^{x}\partial
_{3}^{y}\nabla ^{-2} & 0 & 0 & 0 & 0 \\ 
0 & 0 & 0 & 0 & 0 & 0 & 0 & 0 \\ 
0 & 0 & a & b & -2\left( v^{2}\partial _{x}^{3}-v^{3}\partial
_{x}^{2}\right) \nabla ^{-2}\left( x,y\right) & \delta +\partial
_{1}^{x}\partial _{y}^{1}\nabla ^{-2} & \partial _{1}^{x}\partial
_{y}^{2}\nabla ^{-2} & \partial _{1}^{x}\partial _{y}^{3}\nabla ^{-2} \\ 
0 & -a & 0 & c & -2\left( v^{3}\partial _{x}^{1}-v^{1}\partial
_{x}^{3}\right) \nabla ^{-2}\left( x,y\right) & \partial _{2}^{x}\partial
_{y}^{1}\nabla ^{-2} & \delta +\partial _{2}^{x}\partial _{y}^{2}\nabla ^{-2}
& \partial _{2}^{x}\partial _{y}^{3}\nabla ^{-2} \\ 
0 & -b & -c & 0 & -2\left( v^{2}\partial _{x}^{1}-v^{1}\partial
_{x}^{2}\right) \nabla ^{-2}\left( x,y\right) & \partial _{3}^{x}\partial
_{y}^{1}\nabla ^{-2} & \partial _{3}^{x}\partial _{y}^{2}\nabla ^{-2} & 
\delta +\partial _{3}^{x}\partial _{y}^{3}\nabla ^{-2}%
\end{array}
\right]  \notag ~,
\end{equation*}
\vspace{1cm}
\begin{eqnarray*}
a &=&[-v^{3}\left( \partial _{x}^{2}\partial _{2}^{y}+\partial
_{x}^{1}\partial _{1}^{y}\right) +v^{1}\partial _{1}^{x}\partial
_{y}^{3}+v^{2}\partial _{2}^{x}\partial _{y}^{3}]\nabla ^{-2}\left(
x,y\right) ~,  \label{c4} \\
b &=&[v^{2}\left( \partial _{x}^{3}\partial _{3}^{y}+\partial
_{x}^{1}\partial _{1}^{y}\right) -v^{1}\partial _{1}^{x}\partial
_{y}^{2}-v^{3}\partial _{3}^{x}\partial _{y}^{2}]\nabla ^{-2}\left(
x,y\right) ~,  \notag \\
c &=&[-v^{1}\left( \partial _{x}^{2}\partial _{2}^{y}+\partial
_{x}^{3}\partial _{3}^{y}\right) +v^{3}\partial _{3}^{x}\partial
_{y}^{1}+v^{2}\partial _{2}^{x}\partial _{y}^{1}]\nabla ^{-2}\left(
x,y\right) ~.  \notag
\end{eqnarray*}
\end{sidewaystable}
It's a matter of following the prescription \cite{Marco3}

\bigskip 
\begin{equation}
\xi _{\mu }^{\ast }\left( x\right) =\int dy\text{ }\Lambda _{\mu }^{\nu
}(x,y)\text{ }\xi _{\nu }\left( y\right)  \label{L}
\end{equation}

to get the projected coordinates:

\begin{eqnarray}
\xi _{1}^{\ast }\left( x\right) &=&0,  \label{c5} \\
\xi _{2}^{\ast }\left( x\right) &=&A^{1\perp }\left( x\right) ,  \notag \\
\xi _{3}^{\ast }\left( x\right) &=&A^{2\perp }\left( x\right) ,  \notag \\
\xi _{4}^{\ast }\left( x\right) &=&A^{3\perp }\left( x\right) ,  \notag \\
\xi _{5}^{\ast }\left( x\right) &=&0,  \notag \\
\xi _{6}^{\ast }\left( x\right) &=&\pi _{1}^{\perp }\left( x\right)
+aA^{2}+bA^{3},  \notag \\
\xi _{7}^{\ast }\left( x\right) &=&\pi _{2}^{\perp }\left( x\right)
-aA^{1}+cA^{3},  \notag \\
\xi _{8}^{\ast }\left( x\right) &=&\pi _{3}^{\perp }\left( x\right)
-bA^{1}-cA^{2}.  \notag
\end{eqnarray}%
Due to the presence of transverse fields, \ we simplify, without any lost of
generality, the set of coordinates by choosing a direction of propagation.
So, making $\mathbf{k}=\left( 0,0,k\right) $ we get the simpler set: 
\begin{eqnarray}
\xi _{2}^{\ast } &=&A^{1}\rightarrow q_{1},  \label{c6} \\
\xi _{3}^{\ast } &=&A^{2}\rightarrow q_{2},  \notag \\
\xi _{6}^{\ast } &=&\pi _{1}+v^{2}A^{3}\rightarrow p_{1},  \notag \\
\xi _{7}^{\ast } &=&\pi _{2}-v^{1}A^{3}\rightarrow p_{2},  \notag \\
\xi _{8}^{\ast } &=&-v^{2}\xi _{2}^{\ast }+v^{1}\xi _{3}^{\ast }\rightarrow
-v^{2}q_{1}+v^{1}q_{2}.  \notag
\end{eqnarray}%
The notation (q,p) turns it more evident that we are in 4-dimensional phase
space.

The canonical Hamiltonian, written with symplectic coordinates is 
\begin{eqnarray}
\mathcal{H}\left( \xi \right)  &=&\frac{1}{2}\xi _{i+5}\xi _{i+5}+\frac{1}{2}%
\varepsilon _{ijk}\partial ^{i}\xi _{j+1}\varepsilon _{lmk}\partial ^{l}\xi
_{m+1}+\frac{1}{2}\varepsilon _{0ijk}\left( \xi _{i+5}v^{j}\xi
_{k+1}+v^{0}\varepsilon _{0ijk}\xi _{k+1}\partial ^{i}\xi _{j+1}\right) +
\label{Ham2} \\
&&-\frac{1}{8}v_{j}(v^{j}\xi _{k+1}\xi _{k+1}+\xi _{j+1}v_{k}\xi _{k+1}), 
\notag
\end{eqnarray}%
so that the physical Hamiltonian is 
\begin{eqnarray}
\mathcal{H}^{\ast }\left( \xi ^{\ast }\right)  &=&\frac{1}{2}\xi
_{i+5}^{\ast }\xi _{i+5}^{\ast }+\frac{1}{2}\varepsilon _{ijk}\partial
^{i}\xi _{j+1}^{\ast }\varepsilon _{lmk}\partial ^{l}\xi _{m+1}^{\ast }+%
\frac{1}{2}\varepsilon _{0ijk}\left( \xi _{i+5}^{\ast }v^{j}\xi _{k+1}^{\ast
}+v^{0}\xi _{k+1}^{\ast }\partial ^{i}\xi _{j+1}^{\ast }\right) +
\label{Ham3} \\
&&-\frac{1}{8}v_{j}\left( v^{j}\xi _{k+1}^{\ast }\xi _{k+1}^{\ast }+\xi
_{j+1}^{\ast }v_{k}\xi _{k+1}^{\ast }\right).   \notag
\end{eqnarray}%
Written in $(q,p)$ notations it is: 
\begin{eqnarray}
\mathcal{H}^{\ast }\left( q,p\right)  &=&\frac{1}{2}\left(
p_{1}^{2}+p_{2}^{2}\right) +\frac{1}{2}\left( \partial ^{3}q_{2}\right) ^{2}+%
\frac{1}{2}\left( \partial ^{3}q_{1}\right) ^{2}+  \label{Ham4} \\
&&-\frac{1}{2}v^{3}\left( p_{1}q_{2}-p_{2}q_{1}\right) -\frac{1}{4}\beta
\left( q_{1}^{2}+q_{2}^{2}\right) -\frac{1}{2}v_{0}\left( q_{1}\partial
^{3}q_{2}-q_{2}\partial ^{3}q_{1}\right),   \notag
\end{eqnarray}%
\begin{equation}
\beta \equiv v_{1}^{2}+v_{2}^{2}+\frac{1}{2}v_{3}^{2}.  \label{c7}
\end{equation}%
The equations of motion are thus: 
\begin{eqnarray}
\overset{\cdot }{q}_{1} &=&p_{1}-\frac{1}{2}v_{3}q_{2},  \label{c8} \\
\overset{\cdot }{q}_{2} &=&p_{2}+\frac{1}{2}v_{3}q_{1},  \notag \\
\overset{\cdot }{p}_{1} &=&\frac{1}{2}v_{3}p_{2}-\frac{1}{2}\beta
q_{1}-\partial ^{3}\partial ^{3}q_{1}-v_{0}\partial ^{3}q_{2},  \notag \\
\overset{\cdot }{p}_{2} &=&-\frac{1}{2}v_{3}p_{1}-\frac{1}{2}\beta
q_{2}-\partial ^{3}\partial ^{3}q_{2}+v_{0}\partial ^{3}q_{1},  \notag
\end{eqnarray}%
or, 
\begin{eqnarray}
\square q_{1} &=&\frac{1}{2}v_{j}v^{j}q_{1}-v_{0}\partial ^{3}q_{2},
\label{c9} \\
\square q_{2} &=&\frac{1}{2}v_{j}v^{j}q_{2}+v_{0}\partial ^{3}q_{1}.  \notag
\end{eqnarray}%
To solve this system we decouple the equations and we obtain: 
\begin{equation}
\left[ \left( \square -\frac{1}{2}v_{j}v^{j}\right) ^{2}+v_{0}^{2}(\partial
^{3})^{2}\right] q_{1}=0.  \label{J}
\end{equation}%
The dispersion relation obtained is 
\begin{equation}
\left( k^{\alpha }k_{\alpha }-\frac{1}{2}v_{j}v^{j}\right)
^{2}+v_{0}^{2}(ik^{3})^{2}=0.  \label{J0}
\end{equation}%
We would like to compare this equation with the dispersion relation of the
work \cite{Jackiw} 
\begin{equation}
\left( k^{\alpha }k_{\alpha }\right) ^{2}+\left( k^{\alpha }k_{\alpha
}\right) \left( v^{\beta }v_{\beta }\right) =\left( k^{\alpha }v_{\alpha
}\right) ^{2}.  \label{J1}
\end{equation}%
To this end we rewrite it, first, with the prescription $k^{\mu }=\left(
k^{0};0,0,k^{3}\right) $ and $v^{\mu }=(0;0,0,v^{3});$ we have: 
\begin{equation}
(k_{0})^{2}=\frac{(2(k_{3})^{2}+(v_{3})^{2})\pm v_{3}\sqrt{%
(v_{3})^{2}+4(k_{3})^{2}}}{2}.  \label{spa1}
\end{equation}%
This choice of $v^{\mu }$ is convenient because we can avoid unstable
solutions. In our case we have, from eq. \ref{J0} 
\begin{equation}
(k_{0})^{2}=(k_{3})^{2}+\frac{1}{2}(v_{3})^{2}.  \label{spa2}
\end{equation}

In both cases we have stable solutions. The configuration of a $v^{\mu }$
space-like do not generate imaginary frequencies.

On the other hand, if we adopt the prescription: $k^{\mu }=\left(
k^{0};0,0,k^{3}\right) $ and $v^{\mu }=(v^{0};0,0,0),$ to eq. \ref{J1}\ we
have: 
\begin{equation}
(k_{0})^{2}=(k_{3})^{2}\pm \left\vert v_{0}k_{3}\right\vert ,  \label{time1}
\end{equation}%
and in our case, eq. \ref{J0}, we have 
\begin{equation}
(k_{0})^{2}=(k_{3})^{2}\pm \left\vert v_{0}k_{3}\right\vert.  \label{time2}
\end{equation}%
With this comparison we can extract similar conclusion which has been
discussed about this type of Lorentz breaking, i. e., that we must take
careful with the $v_{0}$ component. If we observe the last solution above we
conclude that we can obtain imaginary frequencies. This situation
characterizes tachionic modes and certainly we have problems with the
causality of this model. On the other hand, the conclusions about the
results in both dispersion relations (\ref{spa1},\ref{spa2}) with $v^{\mu }$
space-like\ is similar. In this case we don't have problems with imaginary
frequencies and the causality in this case is completely assured. Therefore,
by the analysis using SPM, we can arrive to the results about this type of
Lorentz breaking in the recent literature: the configuration of $v^{\mu }$
space-like is stable, and the $v^{\mu }$ time-like carry tachionic \ modes.%
\emph{\ }

\section{Concluding Comments}

The central motivation of this work was to revisit the model of Lorentz and
CPT violation proposed by Jackiw et. al \cite{Jackiw} in the context of the
canonical formalism, since, as \ far we know, it had not been worked out up
to present days. To this end, we applied the recently developed SPM, so that
some relevant physical aspects could be easily read.

The first interesting point to remark is that the conditions on the vector $%
v_{\mu }$, imposed by \cite{Jackiw} after an analysis in order to preserve
gauge invariance, here turns out to be a condition that comes up naturally
still in the context of Dirac's algorithm: we can't get a consistent
constrained theory unless condition (\ref{6}) is obeyed.

After processed the physical variables of the model we got the dispersion
relations to be checked with that of \ \cite{Jackiw}. Using SPM, we arrive
to results that agree with those in the literature: the configuration of $%
v^{\mu }$ space-like is stable, and the $v^{\mu}$ time-like carry tachionic
modes. An interesting discussion that emerge of this analysis is about the
investigations of this scenarios with spontaneous symmetry breaking of gauge
fields. In a Maxwell Electrodynamic we have a wave propagation with only two
transverse modes moving with the same velocity $c$\ (two degrees of
freedom). When we gain a mass term in the Lagrangian \ref{action1}, provided
by spontaneous symmetry breaking, the longitudinal mode can not be
eliminated. Therefore we gain plus one degree of freedom. We have an
interaction mediating massive fotons (short range), and we have a screening
of the electromagnetic radiation.

In Maxwell-Chern-Simons model $(1+2$\ $d)$\ \ \cite{Ion} we have only one d.
f. .What's going on in our model? We do not have a massive term coming from
the Higss Mechanism. Therefore we should expect to have a model with only
transversal modes. In fact, the two transverse fields, $q_{1}$ and $q_{2}$
in the eq. \ref{c6} are the physical variables we had got.

The scenarios with spontaneous symmetry breaking is under investigation and
we shall soon be reporting on it in \cite{novo}.

\section{ Acknowledgment}

The authors are very grateful to J. A. Helayel Neto and I. V. Vancea for very clarifying
discussions. One of the authors (H. Belich) expresses his gratitude to CNPq
for the invaluable financial help.


\begin{thebibliography}{99}
\bibitem{observ} D. Colladay and V. A. Kosteleck\'{y}, \emph{Phys. Rev.}%
\textit{{\ \textbf{D}}} \textbf{55},6760 (1997);

\bibitem{Jackiw} S. Carroll, G. Field and R. Jackiw, \emph{Phys. Rev.} 
\textbf{D 41}, 1231 (1990);

\bibitem{Kostelec1} V. A. Kostelecky and S. Samuel, \emph{Phys. Rev.} 
\textbf{D}$,$\textbf{39}, 683 (1989); V. A. Kostelecky and R. Potting, \emph{%
Nucl. Phys.} \textbf{B359}, 545 (1991); ibid, \emph{Phys. Lett.}\textbf{B381}%
, 89 (1996); V. A. Kostelecky and R. Potting, \emph{Phys. Rev}. \textbf{D51}%
, 3923 (1995);

\bibitem{Colladay} D. Colladay and V. A. Kosteleck\'{y}, \emph{Phys. Rev.}%
\textit{{\ \textbf{D }}}\textbf{58}, 116002 (1998); S.R. Coleman and S.L.
Glashow, \emph{Phys. Rev.} \textbf{D 59}, 116008 (1999).

\bibitem{Berger} M. S. Berger and V. A. Kosteleck\'{y}, \emph{Phys.Rev.} 
\textbf{D 65}, 091701 (2002); M.S. Berger, \textit{Phys. Rev. }\textbf{D 68}%
\textit{, }115005 (2003).

\bibitem{nos} \ H. Belich et al., \emph{Phys.Rev}. \textbf{D 68} (2003)
065030; \emph{Nucl. Phys}\textit{{. \textbf{B}}.- Supp. }\textbf{127, }105
-109, (2004)

\bibitem{NIBBELINK} \ S. G. Nibbelink, M. Pospelov hep-ph/0404271 (2004).

\bibitem{Cosmic} J. W. Moffat, Int. J. Mod. Phys. \textbf{D12}, 1279 (2003);
O. Bertolami, hep-ph/0301191.

\bibitem{Coleman} S.R. Coleman and S.L. Glashow, Phys. Rev. \textbf{D 59},
116008 (1999); V. A. Kosteleck\'{y} and M. Mewes, Phys. Rev. Lett. \textbf{%
87 }, 251304 (2001); Phys. Rev. \textbf{D 66}, 056005 (2002).

\bibitem{Particles} E. O. Iltan, Mod. Phys. Lett. \textbf{A19}, 327 (2004);
-ibid, hep-ph/0308151.

\bibitem{Coupling} A. Kostelecky, R. Lehnert and M. J. Perry, Phys. Rev. 
\textbf{D 68}, 123511 (2003), O. Bertolami \textit{et. al.}, Phys. Rev. 
\textbf{D} \textbf{69}, 083513 (2004).

\bibitem{Astrophys} M. Goldhaber and V. Timble, \textit{J. Astrophys.
Astron. } \textbf{17}, 17 (1996); D. Hutsem\'{e}kers and H. Lamy, \textit{%
Astron. Astrophys. }\textbf{332}, 410 (1998); D. Hutsem\'{e}kers and H.
Lamy, \textit{Astron. Astrophys. }\textbf{367}, 381 (2001); V. B. Bezerra,
H. J. Mosquera Cuesta, C. N. Ferreira \emph{Phys. Rev}. \textbf{D67}, 084011
(2003).

\bibitem{colatto} L. P. Colatto, A. L. A. Penna, and W. C. Santos;
hep-th/0310220, to appear in Eur. Phys. Jour. C.

\bibitem{Adam} C. Adam and F. R. Klinkhamer, \emph{Nucl. Phys}\textit{{. 
\textbf{B}}} \textbf{607}, 247 (2001); C. Adam and F.R. Klinkhamer, \emph{%
Phys. Lett}. \textbf{B }513, 245 (2001) ; V.A. Kostelecky and R. Lehnert, 
\emph{Phys. Rev}. \textbf{D 63,} 065008 (2001) ; A.A. Andrianov, P. Giacconi
and R. Soldati, JHEP 0202, 030 (2002); A.A. Andrianov, R. Soldati and L.
Sorbo, \emph{Phys. Rev}\textit{.} \textbf{D 59}, 025002 (1999).

\bibitem{Chung} R. Jackiw and V. A. Kosteleck\'{y}, \emph{Phys. Rev. Lett.} 
\textbf{82}, 3572 (1999); J. M. Chung and B. K. Chung \emph{Phys. Rev}%
\textit{. }\textbf{D} \textbf{63}, 105015 (2001); J.M. Chung, Phys.Rev. 
\textbf{D 60}, 127901 (1999); M. Perez-Victoria, Phys. Rev. Lett. \textbf{83}%
, 2518 (1999); Bonneau,\emph{\ Nucl.Phys.} \textbf{B 593, }398 (2001) ; M.
Perez-Victoria, \emph{JHEP} 0104, 032 (2001).

\bibitem{Belich} A. P. Ba\^{e}ta Scarpelli, H. Belich, J. L. Boldo and J. A.
Helay\"{e}l-Neto, \emph{Phys. Rev. }\textbf{D} \textbf{67}, 085021 (2003).

\bibitem{Manojr1} H. Belich, M.M. Ferreira Jr., J. A. Helay\"{e}l-Neto and
M.T.D. Orlando, \emph{Phys. Rev. }\textbf{D} \textbf{67}, 125011 (2003).

\bibitem{Manojr2} H. Belich, M.M. Ferreira Jr., J. A. Helay\"{e}l-Neto and
M.T.D. Orlando, \emph{Phys. Rev. }\textbf{D} \textbf{68}, 025005 (2003)

\bibitem{BRST} M. Henneaux, C. Teitelboim, \emph{Quantization of Gauge
Systems}, (Princeton University Press, 1992)

\bibitem{Amaral1} C.~M.~ Amaral, Nuovo Cim.\ B \textbf{25}, 817 (1975).

\bibitem{Amaral2} P.~Pitanga and C.~M.~ Amaral, Nuovo Cim.\ A \textbf{103}
(1990) 1529.

\bibitem{Marco1} M.~A.~ Santos, J.~C.~de Mello and F.~R.~Simao, J.\ Phys.\ A 
\textbf{21}, L193 (1988).

\bibitem{Marco2} M.~A.~Santos, J.~C.~de Mello and P.~Pitanga, Z.\ Phys.\ C 
\textbf{55}, 271 (1992).

\bibitem{Marco3} M.~A.~De Andrade, M.~A.~Santos and I.~V.~Vancea, Mod.\
Phys.\ Lett.\ A \textbf{16}, 1907 (2001)[arXiv:hep-th/0108197].

\bibitem{Marcotese} M.~A.~Dos Santos, hep-th/0202078.

\bibitem{Marco4} M.~A.~De Andrade, M.~A.~Santos and I.~V.~Vancea,
arXiv:hep-th/0308169.

\bibitem{Marco5} M.~A.~De Andrade, M.~A.~Santos and I.~V.~Vancea, JHEP 
\textbf{0106}, 026 (2001) [arXiv:hep-th/0104154].

\bibitem{Marco6} L.~R.~Manssur, A.~L.~Nogueira and M.~A.~Santos, Int.\ J.\
Mod.\ Phys.\ A \textbf{17}, 1919 (2002) hep-th/0005214. In arbirary
dimension:

\bibitem{Bekaert1} X.~Bekaert and A.~Gomberoff, JHEP \textbf{0301}, 054
(2003) [arXiv:hep-th/0212099].

\bibitem{Hong1} S.~T.~Hong, Y.~W.~Kim, Y.~J.~Park and K.~D.~Rothe, J.\
Phys.\ A \textbf{35}, 7461 (2002) [arXiv:hep-th/0204188].

\bibitem{Ghosh1} S.~Ghosh, Annals Phys.\ \textbf{291}, 1 (2001)
[arXiv:hep-th/0009165].

\bibitem{Leal1} L.~Leal and O.~Zapata, Phys.\ Rev.\ D \textbf{63}, 065010
(2001) [arXiv:hep-th/0008049].

\bibitem{Fleck1} M.~Fleck, arXiv:hep-th/0007098.

\bibitem{Haller1} K.~Haller and E.~Lim-Lombridas, Annals Phys.\ \textbf{246}%
, 1 (1996) [Erratum-ibid.\ \textbf{257}, 205 (1997)] [arXiv:hep-th/9409133].

\bibitem{Devecchi1} F.~P.~Devecchi, M.~Fleck, H.~O.~Girotti, M.~Gomes and
A.~J.~da Silva, Annals Phys.\ \textbf{242}, 275(1995) [arXiv:hep-th/9411224].

\bibitem{Lee1} H.~l.~Lee, Y.~W.~Kim and Y.~J.~Park, J.\ Korean Phys.\ Soc.\ 
\textbf{30}, 23 (1997) [arXiv:hep-th/9711186].

\bibitem{Girotti1} H.~O.~Girotti, Int.\ J.\ Mod.\ Phys.\ A \textbf{14}, 2495
(1999) [arXiv:hep-th/9803153].

\bibitem{Li1} S.~Li and Y.~s.~Duan, Astrophys.\ Space Sci.\ \textbf{268},
455 (1999) [arXiv:gr-qc/9808078].

\bibitem{Ghosh2} S.~Ghosh, arXiv:hep-th/9901051.

\bibitem{Itoh1} T.~Itoh and P.~Oh, Phys.\ Rev.\ D \textbf{63}, 025019 (2001)
[arXiv:hep-th/0006163].

\bibitem{Ion} J.A. Helayel-Neto, M.A. Santos, I.V. Vancea,
arXiv:hep-th/0407146.

\bibitem{novo} H. Belich, M. A. Santos, work in progress.
\end{thebibliography}
\end{document}